\documentclass[aps,10pt,nofootinbib]{revtex4}
\usepackage{amsmath,amssymb,amsfonts}
\usepackage{epsfig,graphicx}

\begin{document}
\title{Renormalization group analysis of cosmological constraint
on the mass of Higgs scalar}
\author{\textbf{V. V. Kiselev}$^{1),2)}$, \textbf{S. A. Timofeev}$^{1),2)}$}
\email[E-mail address: ]{serg_timofeev@mail.ru}
\affiliation{$^{1)}$Institute for High Energy Physics, Protvino, Russia\\
$^{2)}$Moscow Institute of Physics and Technology, Dolgoprudny, Russia}
\begin{abstract}
The Higgs boson of Standard Model, minimally coupled to the gravitation, is
not able to produce the inflation of early Universe if its mass exceeds the
threshold value, which is equal to $m_\mathrm {H}^\mathrm{min}=142$~GeV in
the tree approximation for the scalar potential. Two-loop corrections modify
the estimate as $m_\mathrm{H}^\mathrm{min}=150\pm 3$~GeV, so that
higher-order corrections of perturbation theory are completely under control,
though they are numerically important in respect of experimental searches.
\end{abstract}
\maketitle

\section{Introduction}

The stage of inflationary expansion of Universe was recently accepted as the
reference model for the evolution of early Universe
\cite{infl,i-Linde,i-Albrecht+Steinhardt,i-Linde2,inflation}. In
\cite{cosmo-constraint} we have considered the properties of inflation,
produced by the scalar Higgs boson of Standard model with the minimal
coupling to the gravitation, i.e., when the term of lagragian in the form of
$\xi \Phi^\dagger\Phi\, R$ gets zero constant $\xi=0$ (here ${\Phi}$ is the
Higgs field, $R$ is the scalar curvature). Since the vacuum expectattion
value of Higgs field is negligible with respect to the Planck scale of
energy, characteristic for the inflation regime, one can neglect terms
quadratic to the field in potential $V(\Phi)$, so that the potential is
reduced to the form of $V=\lambda (\Phi^\dagger\Phi)^2$. The inflation with
the quartic self-action of inflaton was in detail studied in the framework of
slowly drifting stable attractor appearing in the system of field equations
to the leading approximation of flat homogeneous and isotropic Universe
\cite{KT3,KT-GERG}. In \cite{cosmo-constraint} we have shown that exclusion
of inflation produced by the Higgs boson results in the constraint on the
constant $\lambda$: $\lambda>\frac16$, that leads to a critical minimal value
of boson mass. This constraint is related with the fact that the inflation
produced by the Higgs field is finished at the Hubble constant $H$
determining the rate of Universe expansion as given by the formula
\cite{cosmo-constraint}
\begin{equation}\label{inf-end}
    2\pi G\,H^2=\lambda,
\end{equation}
where $G$ is the gravitational constant. The inflation cannot be produced by
the Higgs field, if the constant of self-action is close to unit, hence, the
Hubble constant to the end of inflation would be of the order of Planck
energy. But the Planck scales of energy density cannot be described in the
framework of classical theory of gravitation, which is the necessary
ingredient of inflationary model. The numerical consideration of scheme for
the mentioned derivation of decoupling constant of self-action for the Higgs
scalar gives the critical value of $\lambda_c=\frac16$ by defining a limit of
quantum gravity in cosmology. Then, the mass of Higgs field should exceed the
decoupling value\footnote{We suggest here that the Higgs boson is the only
scalar field in the theory, so that the Higgsian inflation is not consistent
with observations, while the only possible variant is the ordinary Big Bang
due to oscillations of Higgs field in vicinity of minimum of its potential
with a fine tuning of initial data consistent with the large scale structure
of Universe. The case with introducing the additional scalar field
responsible of the inflation  and two field dynamics in early Universe will
be considered elsewhere.} equal to $m_\mathrm{H}^\mathrm{min}=142$ GeV to the
tree approximation for the potential. Then, after the determination of
decoupling mass to the leading order, one has got the problem to take into
account loop corrections. These corrections depend on the energy scale,
corresponding to the phenomenon under consideration, therefore we use the
renormalization group up to two loops in order to account for the higher
corrections of perturbation theory. First, we estimate the energy scale
relevant to the end of inflation. Second, we study the dependence of final
result on the initial data and scale variations fixing the running constants
and masses.

\section{Estimating the energy scale}
We can estimate the characteristic scale of energy in the threshold region of
inflationary regime in several methods.
\subsection{Characteristic value of field}
Let the running constant of self-action be given at the scale fixed by the
field value $\lambda=\lambda(\phi)$, where the real electrically neutral
field $\phi$ is given by the gauge $\phi=\Phi\sqrt2$. Then, we set on
$\lambda=\frac16$ in the relation for the threshold value of Hubble constant
in (\ref{inf-end}), and we make use of Einstein equations taking into account
the fact that at the threshold the kinetic energy is twice the potential one
\cite{cosmo-constraint}, hence,
\begin{equation}
H^2 = 2\pi G \lambda \phi^4,
\end{equation}
we find at $\mu=\phi$:
\begin{equation}\label{1}
\mu = \sqrt\frac{1}{2\pi G}\approx 4.9 \times 10^{18} \mbox{ GeV}.
\end{equation}

\subsection{The energy density}
The scale of energy can be estimated by the value of energy density by
$\rho=\mu^4$, so that due to the Einstein equations
\begin{equation}
H^2=\frac{8\pi G}{3}\rho,
\end{equation}
we get
\begin{equation}\label{2}
\mu=\sqrt\frac{1}{4\pi G \sqrt{2}}\approx 2.9 \times 10^{18}
\mbox{ GeV},
\end{equation}
that very slightly differs from the result in (\ref{1}).

\subsection{The field virtuality}

The displacement of virtual field from the mass shell gives $\mu^2=m^2-p^2$,
where the mass equals to
$$m^2=\frac{\partial ^2 V}{\partial \phi ^2}=3\lambda \phi^2=\frac{1}{4
\pi G},$$ while the 4-momentum $p$ is estimated by $$p_0 \phi = i
\frac{\partial \phi}{\partial t},$$ whereas $\boldsymbol{\mathrm{p}}=0$,
since the Higgs field is spatially homogeneous and isotropic in the reference
system under consideration. Finally, we get
$$p_0^2=-\frac{(\dot{\phi})^2}{\phi^2}=-\frac{1}{12\pi G}.$$
Here we have used the values of $\phi$ and $\dot{\phi}$, as derived in
\cite{cosmo-constraint}. Thus, we find
\begin{equation}
\mu^2=m^2-p^2=\frac1{3\pi G},
\end{equation}
\begin{equation}
\mu=\sqrt\frac1{3\pi G}\approx 4.0 \times 10^{18} \mbox{ GeV}.
\end{equation}

Therefore, we conclude that the energy scale has got the Planckian order, so
that it is estimated as $3\times10^{18}$ GeV.

\section{The renormalization group analysis}
According to \cite{Amsler:2008zzb}, the experimental values of masses and
coupling constants entering as initial data into the renormalization group
equations are equal to
\begin{eqnarray}
\nonumber m_Z &=&91.1873\pm 0.0021\mbox{ GeV},\\
\nonumber m_t &=&170.9\pm 1.9\mbox{ GeV},\\
\alpha^{-1}_\mathrm{em}(m_Z)&=&127.906 \pm 0.019,\\
\nonumber \alpha_s(m_Z)&=&0.1187\pm 0.0020,\\
\nonumber \sin^2\theta_\mathrm{W}&=&0.2312\pm 0.002,
\end{eqnarray}
wherein the running constants are normalized at the scale equal to the mass
of $Z$-boson, while the sine of Weinberg angle is standardly defined  in
terms of renormalized values of coupling constants in the electroweak group,
$g$ and $g'$.

It is convenient to take those quantities at scale $\mu=m_t$, so that
\begin{eqnarray}
\nonumber g^\prime&=&0.358765\pm 0.00010,\\
g&=&0.648532\pm 0.00039,\\
\nonumber g_s&=&1.17372\pm0.0099.
\end{eqnarray}

The relation between the running constant $\lambda(\mu)$ and Yukawa constant
$h_t(\mu)$ for the $t$-quark at scale $\mu=m_t$ is given in terms of Higgs
scalar mass and $t$-quark mass, $m_t$, as
\begin{equation}
\lambda (m_t)=\frac{m_\mathrm{H}^2}{2v^2} (1+\Delta_\mathrm{H}),
\end{equation}
\begin{equation}
h_t (m_t)=\frac{\sqrt{2}}{v}m_t (1+\Delta_t),
\end{equation}
where the vacuum expectation value of Higgs field $v$ = 246.2 GeV is related
to the Fermi constant of weak interaction, while corrections
$\Delta_{\mathrm{H},t}$ are given in Appendix 1 up to the 1-loop accuracy
(formulas of calculations refer to the scheme $\overline{\mathrm{MS}}$ and
they are taken from \cite{Espinosa:2007qp}.).

The renormalization group equations (see Appendix 2 extracted from
\cite{Espinosa:2007qp}) show that the critical value of Higgs boson mass is
displaced from its tree level value by 10~GeV upper in the 1-loop
approximation and by 8~GeV upper in 2 loops.

We calculate the dependence of result on the initial data such as $\alpha_s$,
$m_t$ and $\mu$, since they involve the main uncertainty into the estimates
of cosmological constraint on the Higs boson mass, as we find numerically.

Then, in 2-loop approximation the dependence of critical value of Higgs
scalar mass $m_\mathrm{H}^\mathrm{min}$ on the parameters can be presented in
terms of partial derivatives, so that
\begin{eqnarray}
\frac{\partial m_\mathrm{H}^\mathrm{min}}{\partial \ln\mu}&=&-0.28 \mbox{ GeV},\\
\label{diff}
\frac{\partial m_\mathrm{H}^\mathrm{min}}{\partial \alpha_{s}(m_t)}&=&-110 \mbox{ GeV},\\
\frac{\partial m_\mathrm{H}^\mathrm{min}}{\partial m_t}&=&1.0.
\end{eqnarray}

We transform the derivative with respect to the strong interaction constant
$\alpha_s$ from scale $m_t$ in (\ref{diff}) to scale $m_Z$. So, we have used
formulas from Appendix 2 for the loop corrections to charges. In 1-loop
approximation we find
\begin{equation}
\frac1{\alpha_s(m_z)}=\frac1{\alpha_s(m_t)}+\frac{\beta_0}{2\pi}\ln\frac{m_Z}{m_t},
\end{equation}
where $\beta_0=11-\frac23 n_f$, and $n_f$ is the number of active quark
flavors at scales $m_Z<\mu<m_t$, i.e. $n_f=5$. Thus,
\begin{equation}
\frac{\partial m_\mathrm{H}^\mathrm{min}}{\partial \alpha_{s}(m_z)}=-93.8 \mbox{
GeV}.
\end{equation}

Therefore, the results can be presented in the form
\begin{equation}\label{fin}
m_\mathrm{H}^\mathrm{min}= 150+0.28\ln\frac{10^{18}}{\mu}-0.19
\frac{\alpha_s-0.1187}{0.002}+2\frac{m_t-171}{2} \pm
2\mbox{ GeV},
\end{equation}
where $\mu$ and $m_t$ are expressed in GeV.

We see from (\ref{fin}) that the dependence on the energy scale is weak,
therefore, the order of magnitude for the scale is important, only, as we
have expected above. The strong dependence on the mass of $t$-quark is
evident. Further, the difference between the 1- and 2-loop results is about 2
GeV, so we can conservatively prescribe its value to the uncertainty due to
higher orders of perturbation theory in the framework of renormalization
group as shown in (\ref{fin}).

\section{Discussion and Conclusion}
We have calculated the critical value of Higgs boson mass
$m_\mathrm{H}^\mathrm{min}$ in the Standard model minimally coupled to the
gravitation up to 2-loop corrections in the effective action. The critical
mass determines the cosmological constrain. In the framework of
renormalization group we have estimated variations of
$m_\mathrm{H}^\mathrm{min}$ with respect to small changes of initial data
related to the uncertainties of experimental measurements as well as at
different prescriptions for the energy scale characteristic for the final
stage of Universe inflation, which can be produced by the Higgs field. The
obtained result is rather stable with respect to higher corrections of
perturbation theory, so that the uncertainty of calculations due to this
factor gives the value of 2 GeV. The other significant source of uncertainty
of calculations is the mass of $t$-quark. Finally, we deduce the value of
decoupling mass from cosmology as ${m_\mathrm{H}^\mathrm{min}=150\pm3}$~GeV.
In the Standard model there is the upper limit on the Higgs boson mass
$m_\mathrm{H}<189$ GeV, appearing from the following requirement: the Landau
pole in the constant of self-action for the field should be posed at the
scale higher than the Planck mass. Some cosmological consequences because of
the Higgs scalar participation in the inflation, were considered in
\cite{Espinosa:2007qp} with account of thermal and quantum fluctuations of
Higgs field.

Our result should be compared with the model, wherein the Higgs boson of
Standard model is coupled to the gravitation due to the nonminimal
interaction with the constant of $\xi\sim 10^4$
\cite{Barvinsky,Bezrukov,DeSimone,Barvinsky2,Bezrukov2,Burgess,Barbon,
CervantesD,CervantesD-GUT,Barvinsky3}. In that case a conformal
transformation allows one to express the Higgs scalar in terms of new
effective scalar field minimally coupled to the gravitation, so that an
effective potential includes a plateau with a scale of energy, which is
$\sqrt{\xi}$ times less than the Planck mass. Then the inflation becomes
admissible due to the effective field, which parameters are in a consistent
agreement with observed data in cosmology, if the mass of such the Higgs
scalar is constrained within the interval $135.6$ GeV $<m_\mathrm{H}< 184.5$
GeV (see details in \cite{Barvinsky3}). The bound on the Higgs boson mass
derived in our paper in the framework of minimal coupling to the gravitation,
is greater than the lower limit for the case of nonminimal interaction, while
the upper limits are similar in both cases.

This work was partially supported by grants of Russian Foundations for Basic
Research 09-01-12123 and 10-02-00061, Special Federal Program ``Scientific
and academics personnel'' grant for the Scientific and Educational Center
2009-1.1-125-055-008, ant the work of T.S.A. was supported by the Russian
President grant MK-406.2010.2 and the grant of ``The Foundation for the
Support of National Science''.

\appendix

\section*{Appendix 1}
Corrections to the constant of Higgs self-action and Yukawa constant are
given by
\begin{multline}
\Delta_t= - \frac{4\alpha_s(m_t)}{3\pi}+(1.0414N_f -
14.3323)\left[\frac{\alpha_s(m_t)}{\pi}\right]^2 - \frac{4\alpha (m_t)}{9\pi}
+ \\ + \frac{h^2_t}{32\pi^2}\left[\frac{11}2 - r + 2r(2r - 3) \ln(4r) - 8r^2
\left(\frac1r - 1\right)^{3/2} \arccos\sqrt{r}\right] -\\- 6.90 \times
10^{-3} + 1.73 \times 10^{-3} \ln \frac{m_\mathrm{H}}{300\mathrm{GeV}} -
5.82\times 10^{-3} \ln \frac{m_t}{175\mathrm{GeV}}, \tag{A.1}
\end{multline}
where $r=\frac{m_\mathrm{H}^2}{4m_t^2}$, $\alpha_s=g_s^2/(4\pi)$, and
$N_f=5$,
\begin{equation}
\Delta_\mathrm{H} = \frac{G_\mathrm{F}}{\sqrt2} \frac{m_z^2}{16\pi^2}\left[\xi
f_1(\xi) + f_0(\xi) + \xi^{-1}f_{-1}(\xi)\right], \tag{A.2}
\end{equation}
where
\begin{equation}
f_1(\xi)=6 \ln \frac{m_t^2}{m_\mathrm{H}^2} + \frac32 \ln\xi - \frac12
Z\left(\frac1\xi\right)- Z\left(\frac{c_w^2}\xi\right)- \ln c_w^2
+ \frac92 \left(\frac{25}9 - \frac{\pi}{\sqrt{3}}\right),\tag{A.3}
\end{equation}
\begin{multline}
f_0(\xi)=  -6 \ln \frac{m_t^2}{m_z^2}\left[1 + 2c_w^2 - 2
\frac{m_t^2}{m_z^2}\right]+ \frac{3c_w^2\xi}{\xi - c_w^2} \ln
\frac{\xi}{c_w^2} + 2Z\left(\frac{1}{\xi}\right) +
4c_w^2Z\left(\frac{c_w^2}{\xi}\right)+\\+ \left(\frac{3c_w^2}{s_w^2} +
12c_w^2\right)\ln c_w^2 - \frac{15}2(1 + 2c_w^2) -
3\frac{m_t^2}{m_z^2}\left(2Z\left(\frac{m_t^2}{m_z^2\xi}\right)+ 4 \ln
\frac{m_t^2}{m_z^2} - 5\right),\tag{A.4}
\end{multline}
\begin{multline}
f_{-1}(\xi)=  6 \ln \frac{m_t^2}{m_z^2} \left[1 + 2c_w^4 - 4
\frac{m_t^4}{m_z^4}\right]- 6Z\left(\frac{1}{\xi}\right)-\\- 12c_w^4
Z\left(\frac{c_w^2}{\xi}\right)- 12c_w^4 \ln c_w^2 + 8(1 + 2c_w^4)+ 24
\frac{m_t^4}{m_z^4} \left[\ln \frac{m_t^2}{m_z^2} - 2 +
Z\left(\frac{m_t^2}{m_z^2\xi}\right)\right],\tag{A.5}
\end{multline}
with notations: $\xi=\frac{m_\mathrm{H}^2}{m_z^2}$, $s_w^2=\sin^2
\theta_\mathrm{W}$, $c_w^2=\cos^2 \theta_\mathrm{W}$, where
$\theta_\mathrm{W}$ is the Weinberg angle,
\begin{eqnarray}
Z(z)=\nonumber\left\{
\begin{array}{ccc}
2A\arctan (1/A)&(z>1/4),\\
A\ln[(1+A)/(1-A)]&(z<1/4),
\end{array}\right. \qquad A=\sqrt{|1-4z|}.
\end{eqnarray}

\section*{Appendix 2}
The 2-loop equations of renormalization group (RG) for charges take the
following form:
\begin{equation}\label{charge}
\frac{d g_i}{d t}=\kappa g_i^3 b_i + \kappa^2 g_i^3 \left(
\sum_{j=1}^3 B_{ij}g_j^2-d^t_i h_t^2 \right),\tag{A.6}
\end{equation}
where $t=\ln \mu$, and $\kappa = 1/(16\pi^2)$, while
\begin{equation}
b=(41/6,\;-19/6,\;-7),\qquad B=\left(
\begin{array}{ccc}
199/18&9/2&44/3\\
3/2&35/6&12\\
11/6&9/2&-26
\end{array}\right), \qquad d^t=(17/6,\;3/2,\;2).\tag{A.7}
\end{equation}
The RG equation for the Yukawa constant is written down as
\begin{equation}
\frac{d h_t}{d t}=\kappa h_t \left( \frac92 h_t^2-\sum_{i=1}^3
c^t_i g_i^2 \right)+ \kappa^2 h_t \left( \sum_{ij} D_{ij}g_i^2
g_j^2+\sum_{i} E_{i}g_i^2 h_t^2+6(\lambda^2-2h_t^4-2\lambda h_t^2)
\right),\tag{A.8}
\end{equation}
where
\begin{equation}
c^t=(17/12,\;9/4,\;8),\qquad D=\left(
\begin{array}{ccc}
1187/216&0&0\\
-3/4&-23/4&0\\
19/9&9&-108
\end{array}\right), \qquad E=(131/16,\;225/16,\;36).\tag{A.9}
\end{equation}
The RG equation for the cosntant of self-action of Higgs boson reads off
\begin{multline}
\frac{d \lambda}{d t}=\kappa \left\{ -6h_t^4+12h_t^2\lambda+\frac38\left[
2g_4+(g^2+g^{\prime 2})^2 \right]-3\lambda(3g^2+g^{\prime 2})+24\lambda^2
\right\}+\\ +\kappa^2 \left\{  30h_t^6-h_t^4\left( 32g_s^2+\frac83 g^{\prime
2} +3\lambda  \right) +h_t^2 \left[ -\frac94 g^4+\frac{21}2 g^2g^{\prime
2}-\frac{19}4 g^{\prime 4} +\right.\right.\\ \left. +\lambda\left(
80g_s^2+\frac{45}2g^2+\frac{85}6g^{\prime 2}-144\lambda \right)\right]
+\frac1{48}\left( 915g^6-289g^4g^{\prime 2}-559g^2g^{\prime 4} -379g^{\prime
6} \right)+\\+ \left. \lambda\left( -\frac{73}8g^4+\frac{39}4 g^2g^{\prime
4}+\frac{629}{24}g^{\prime 4}+108\lambda g^2+36\lambda g^{\prime
2}-312\lambda^2 \right) \right\}.\tag{A.10}
\end{multline}


\begin{thebibliography}{}

\bibitem{infl}
  A.~H.~Guth,
  Phys.\ Rev.\  D {\bf 23}, 347 (1981).
\bibitem{i-Linde}
  A.~D.~Linde,
  Phys.\ Lett.\  B {\bf 108}, 389 (1982).
\bibitem{i-Albrecht+Steinhardt}
  A.~Albrecht and P.~J.~Steinhardt,
  Phys.\ Rev.\ Lett.\  {\bf 48}, 1220 (1982).
\bibitem{i-Linde2}
  A.~D.~Linde,
  Phys.\ Lett.\  B {\bf 129}, 177 (1983).
\bibitem{inflation}
  A.~Linde,
  Lect.\ Notes Phys.\  {\bf 738}, 1 (2008).
\bibitem{cosmo-constraint}
V.~V.~Kiselev and S.~A.~Timofeev,
  arXiv:0906.4191 [gr-qc].
\bibitem{KT3}
  V.~V.~Kiselev and S.~A.~Timofeev,
  arXiv:0801.2453 [gr-qc].
\bibitem{KT-GERG}
  V.~V.~Kiselev and S.~A.~Timofeev,
  arXiv:0905.4353 [gr-qc] (Gen.\ Rel.\ Grav. (2009) [in press]).
\bibitem{Amsler:2008zzb}
  C.~Amsler {\it et al.}  (Particle Data Group),
  Phys.\ Lett.\  B {\bf 667}, 1 (2008).
\bibitem{Espinosa:2007qp}
  J.~R.~Espinosa, G.~F.~Giudice, and A.~Riotto,
  JCAP {\bf 0805}, 002 (2008)
  [arXiv:0710.2484 [hep-ph]].
\bibitem{Barvinsky}
  A.~O.~Barvinsky and A.~Y.~Kamenshchik,
  Phys.\ Lett.\  B {\bf 332}, 270 (1994)
  [gr-qc/9404062].
\bibitem{Bezrukov}
  F.~L.~Bezrukov and M.~Shaposhnikov,
  Phys.\ Lett.\  B {\bf 659}, 703 (2008)
  [arXiv:0710.3755 [hep-th]].
\bibitem{DeSimone}
  A.~De Simone, M.~P.~Hertzberg, and F.~Wilczek,
  arXiv:0812.4946 [hep-ph].
\bibitem{Barvinsky2}
  A.~O.~Barvinsky, A.~Y.~Kamenshchik, C.~Kiefer, {\it et al.},
  arXiv:0904.1698 [hep-ph].
\bibitem{Bezrukov2}
  F.~Bezrukov and M.~Shaposhnikov,
  arXiv:0904.1537 [hep-ph].
\bibitem{Burgess}
  C.~P.~Burgess, H.~M.~Lee, and M.~Trott,
  arXiv:0902.4465 [hep-ph].
\bibitem{Barbon}
  J.~L.~F.~Barbon and J.~R.~Espinosa,
  arXiv:0903.0355 [hep-ph].
\bibitem{CervantesD}
  J.~L.~Cervantes-Cota and H.~Dehnen,
  Nucl.\ Phys.\  B {\bf 442}, 391 (1995)
  [astro-ph/9505069].
\bibitem{CervantesD-GUT}
J.~L.~Cervantes-Cota and H.~Dehnen,
  Phys.\ Rev.\  D {\bf 51}, 395 (1995)
  [astro-ph/9412032].
\bibitem{Barvinsky3}
  A.~O.~Barvinsky, A.~Y.~Kamenshchik, C.~Kiefer, {\it et al.},
  arXiv:0910.1041 [hep-ph].
\end{thebibliography}
\end{document}